\documentclass[compress,final,3p,times,11pt]{elsarticle}
\usepackage{lineno,hyperref}
\modulolinenumbers[5]
\usepackage{natbib}
\usepackage{caption}
\usepackage{sidecap}
\usepackage[english]{babel}
\usepackage{amsfonts}
\usepackage{amssymb}
\usepackage{amsmath}
\usepackage{color,epstopdf}
\usepackage{mathrsfs}
\usepackage{graphicx,color,epstopdf}
\usepackage{float}
\usepackage{subcaption}
\usepackage{threeparttable}
\usepackage{float}
\usepackage{array}
\usepackage{subcaption}

\newtheorem{theorem}{Theorem}

\journal{...}
\begin{document}
\begin{frontmatter}
\title{A Logistic-Harvest Model with Allee Effect under Multiplicative Noise \tnoteref{mytitlenote}}
\author[mymainaddress,mysecondaryaddress]{Almaz Tesfay \corref{mycorrespondingauthor}}
\ead{amutesfay@hust.edu.cn}
\author[mymainaddress,mysecondaryaddress]{Daniel Tesfay}
\ead{dannytesfay@hust.edu.cn}
\author[mythirdaddress]{James Brannan}
\ead{jrbrn@clemson.edu}
\cortext[mycorrespondingauthor]{Corresponding author}
\author[myfourthaddress]{Jinqiao Duan}
\ead{duan@iit.edu}
\address[mymainaddress]{School of Mathematics and Statistics \& Center for Mathematical Sciences,  Huazhong University of Science and Technology, Wuhan 430074, China}
\address[mysecondaryaddress]{Department of Mathematics, Mekelle University, P.O.Box 231, Mekelle, Ethiopia}
\address[mythirdaddress]{Department of Mathematical Sciences, Clemson University,Clemson, South Carolina 29634, USA}
\address[myfourthaddress]{Department of Applied Mathematics, Illinois Institute of Technology, Chicago, IL 60616, USA\\
}
\begin{abstract}
This work is devoted to the study of a stochastic logistic growth model with and without the Allee effect. Such a model describes the evolution of a population under environmental stochastic fluctuations and is in the form of a stochastic differential equation driven by multiplicative Gaussian noise. With the help of the associated Fokker-Planck equation, we analyze the population extinction probability and the probability of reaching a large population size before reaching a small one. We further study the impact of the harvest rate, noise intensity, and the Allee effect on population evolution. The analysis and numerical experiments show that if the noise intensity and harvest rate are small, the population grows exponentially, and upon reaching the carrying capacity, the population size fluctuates around it. In the stochastic logistic-harvest model without the Allee effect, when noise intensity becomes small (or goes to zero),  the stationary probability density becomes more acute and its maximum point approaches one. However, for large noise intensity and harvest rate, the population size fluctuates wildly and does not grow exponentially to the carrying capacity. So as far as biological meanings are concerned, we must catch at small values of noise intensity and harvest rate. Finally, we discuss the biological implications of our results.
\end{abstract}
\end{frontmatter}
\section{Introduction}\label{Sec1}
A group of individuals of the same species living in a limited place is called a population  \cite{panik2014growth}. The dynamical process of population growth and decline is a function of factors that are intrinsic to a population and the environmental conditions.

The well known logistic growth model describes the growth of population, followed by a reduction, and bound by the maximum population size (carrying capacity). This model is a  nonlinear differential equation
 \begin{equation}\label{log}
\frac{dX_t}{dt}=rX_t\left(1-\frac{X_t}{K}\right), \quad X(0)=x_0, \tag{1.1}
 \end{equation}
where $r > 0$ is the growth rate and $X_t$ is the population size at time $t$ and $K$ is the carrying capacity. This model was first introduced by Verhust \cite{F.Braue(2011)}. When $X_t$ is very small, the equation in (\ref{log}) becomes $\frac{dX_t}{dt}=rX_t$ and $\frac{dX_t}{dt}=0$, when $X_t$ nears the carrying capacity $K$.

Equation (\ref{log}) has a unique solution given by
$X_t=\frac{K}{1+Ae^{-rt}},$ where $A=(\frac{K}{x_0}-1)$. The population size attains its maximum when $t\rightarrow\infty$.

Allee effect was studied widely in a biology book \cite{Allee(2008)}. In this book, the authors cited many papers dealing with the Allee effect. Allee \cite{N. Saber(2009)} suggested that per capita birth rate declines at a low population size ( densities). In this case, the population may go to extinction. The logistic growth model with the Allee effect is one of the most important models in mathematical ecology owing to its theoretical and practical significance. An Allee effect shows a non-negative association between reproduction and population size, and survival of individuals. There are two distinct variations of the Allee effect. Namely, strong Allee effect and weak Allee effect. Strong Allee effect introduces a population threshold \cite{{Michael J.Panik(2017)}} that the population must exceed in order to grow, while the weak  Allee effect does not admit any threshold. For more details about this model see \cite{Qingshan Yang(2011),M. Krstic(2010),Michael J.Panik(2017)} and the reference therein.

The classic general logistic growth model with Allee effect \cite{P. Amarasekare} is given by
\begin{equation}\label{Allee}
\frac{dX_t}{dt}=rX_t\left(\frac{X_t}{S}-1\right)\left(1-\frac{X_t}{K}\right), \qquad X(0)=x_0, \tag{1.2}
\end{equation}
where $X_t$ is the population size at time $t$ in a given area or place, $r > 0$ is the population growth rate, $K > 0$ is the carrying capacity, and $S$ refers to the threshold population (Allee threshold) which is the minimum population that is necessary for the species to survive with values $0 < S < K$.  Extinction occurs whenever the population decreases below the Allee threshold value $S$. Here the initial population size $X_0$ must be greater than the threshold value  $S$. Equation (\ref{Allee}) has two stable equilibrium solutions at $X_1(t) = 0$ and  $X_2(t) = K $, and an unstable equilibrium solution at $X_3(t) = S$.

Based on the resources available to the system, the population should reach the carrying capacity $K$. If the initial population is below the critical threshold $S$, then it approaches extinction as time goes on. Thus the threshold population is useful to biologists in order to determine whether a given species should be placed on the endangered list so that the survival of the species will then be given due attention and necessary protection.

Fishing has a lot of benefits to human beings and it has also a great impact on the socio-economic and infrastructure development of a country. For example, it serves as food, generates income, and creates job opportunities. Many scientists  \cite{B. Dubey,P. Bhattacharya,T. Kumar(2010)} devised strategies to prevent the extinction of renewable resources such as fish by harvesting, and they agreed on the importance. See \cite{H. Kinfe(2016),M. Rahmani(2015),Rahmani M.H.DOUST(2015)} for further explanation on harvesting strategies.

The logistic growth model, with and without the Allee effect, and with harvesting has been used to study the fishery farming \cite{{M. Rahmani(2015)}}. Harvesting is an interesting research area in a population study. The most important input for the successful management of harvested populations is a sustainable strategy. Harvesting strategy should not lead to instabilities or extinctions.

In this paper, we focus on proportional harvesting which removes a fixed proportionality of individuals each time $t$ ( year). In other words, if the population increases, the harvested also increases, and if the population decreases the quantity harvested decreases.

Now let us  consider the mathematical model of the relative-rate harvesting on logistic growth model \cite{F.Braue(2011)} in Eq. (\ref{log-h}) and logistic growth with Allee effect Eq. (\ref{Allee-h}), respectively.

\begin{equation}\label{log-h}
\frac{dX_t}{dt}=rX_t\left(1-\frac{X_t}{K}\right)-\lambda X_t, \qquad X(0)=x_0,\tag{1.3}
\end{equation}
and
\begin{equation}\label{Allee-h}
\frac{dX_t}{dt}=rX_t\left(\frac{X_t}{S}-1\right)\left(1-\frac{X_t}{K}\right)-\lambda X_t, \qquad X(0)=x_0,\tag{1.4}
\end{equation}
where again $r$ is the population growth rate, $K > 0$ is the carrying capacity, and $S$ refers to the Allee threshold, $0 < S < K$ and $\lambda$ is harvest rate.

Equilibria points of Eq. (\ref{log-h}) lie at $X_t=0,$ and $X_t=K\left(1-\frac{\lambda}{r}\right)$ for $\lambda <r$. The potential function $ V(x)$ of Eq. (\ref{log-h}) is given by
$V(x) = r\left[-\frac{1}{2}x^2+\frac{1}{3K}x^3\right] + \frac{\lambda}{2}x^2.$
For $\lambda=0$, the function $V(x)$ becomes the potential function of equation (\ref{log}).

Model (\ref{Allee-h}) has equilibria points at $X_t=0,$ and at the solutions of $\lambda=r\left(\frac{X_t}{S}-1\right)\left(1-\frac{X_t}{K}\right)$. The maximum of the parabola is at $X_t = \frac{(S + K)}{2}$, where we have a saddle-node bifurcation at
$ \lambda=\frac{r(K-S)^2}{4SK}.$

The potential function $ V(x)$ of Eq. (\ref{Allee-h}) for $\lambda\neq0$ is given by
$V(x) = r\left[\frac{1}{2}x^2-\frac{(S+K)}{3SK}x^3 + \frac{1}{4SK}x^4\right] +\frac{\lambda}{2}x^2.$
$V(x)$ reduces to the potential function of equation (\ref{Allee})if $\lambda=0$.

Many researchers \cite{H. Kinfe(2016),M. Hasanbulli,M. Kot(2011),Rahmani M.H.DOUST(2015),N.F.Britton} considered the deterministic model of logistic growth with and without Allee effect under harvesting factor and studied the behavior of the deterministic model free of any stochastic element. Even though deterministic models are much easier to analyze than their corresponding stochastic models, they neglect of random influences on the growth process. stochastic differential equations may be regarded as more adequate models for the development of a population. Since random events affect population dynamics.

In our paper, we focus on both deterministic and  stochastic model. Biological populations exhibit some form of stochastic behavior and that environmental noise should thus be an integral component of any dynamic population model \cite{Michael J.Panik(2017)}. Population ecology deals with demographic and environmental stochasticity. In this work, we consider environmental stochasticity.

Several factors affect the environment population resides \cite{X.Mao(2002)}. To model environmental effects, one possibility is to explicitly include additional variables, for example, chemical agents, food supply, rainfall, and average temperature into differential equation (\ref{log-h}) and (\ref{Allee-h}). On the other hand, population systems are often subject to environmental noise. Thus it is important to reveal how the noise affects the population systems.

According to Equation (12.20)  in \cite{Michael J.Panik(2017)}, a stochastic fishing model is given by a stochastic differential equation (SDE)
\begin{equation}\label{gen}
dX_t= (H(X_t)X_t-\lambda X_t ) dt + \epsilon X_t \; dB_t, \qquad X(0)=x_0,\tag{1.5}
\end{equation}
where $H(X_t)$ is natural growth rate of harvested population, and $\lambda, \epsilon$ are constants. The drift coefficient and diffusion coefficient of this SDE are $f(X_t)=H(X_t)X_t-\lambda X_t$ and $ g(X_t)=  \epsilon^2 X_t^2$,  respectively. This stochastic differential equation has a unique solution \cite{Michael J.Panik(2017)}, and the solution is a homogenous diffusion process.

Here, we choose $H(X)=rX(1-\frac{X}{K}) $ and $H(X)=rX(\frac{X}{S}-1)(1-\frac{X}{K})$. Then by Eq. (\ref{gen}) the stochastic version of the logistic-harvest model of (\ref{log-h}) and (\ref{Allee-h}), respectively are
\begin{equation}\label{log-SDE}
dX_t= \left[rX_t\left(1-\frac{X_t}{K}\right)-\lambda X_t\right] +\epsilon X_tdB_t, \qquad X(0)=x_0,\tag{1.6}
\end{equation}
and
\begin{equation}\label{Allee-SDE}
dX_t= \left[rX_t\left(\frac{X_t}{S}-1\right)\left(1-\frac{X_t}{K}\right)-\lambda X_t\right] + \epsilon X_tdB_t, \qquad X(0)=x_0,\tag{1.7}
\end{equation}
where $B_t$ is a one-dimensional Brownian motion and $\epsilon$ is the Gaussian noise intensity with $0 < \epsilon <1.$

The objective of this work is to investigate the behavior of the logistic-harvest without or with Allee effect, driven by multiplicative Gaussian noise. In other words, we will combine the theory of population biology with that of stochastic differential equations. According to Drake and Lodge \cite{Drake.Lodge(2006)}, there are three statistics most commonly used to evaluate the population helpful in studying stochastic population models. These quantities are the extinction probability, the first passage probability, and the mean time to extinction. In our study, we focus on the extinction probability.

In this paper, we first review the deterministic logistic-harvest model with and without the Allee effect, and then we investigate their stochastic counterpart. We further discuss the extinction probability of the stochastic models. To gain some insight into the logistic-harvest mechanism and consequently about the underlying biological phenomenon, we apply the Euler-Maruyama scheme to approximate the sample solution paths of the stochastic logistic-harvesting model. Finally, we present a short discussion on the comparison between the deterministic models and stochastic models as parameter $x_0,$ $\lambda$ and $\epsilon$ vary.

This paper is arranged as follows: After recalling basic facts about Brownian motion and stochastic differential equations in section $\ref{Sec2a}$, we review and discuss the behavior of the equilibrium solution of the deterministic of the logistic-harvest model without the Allee effect (\ref{log-h}) and analyze  its corresponding stochastic model (section $\ref{Sec3}$). We drive the exact solution of model (\ref{log-SDE}) and explain the effect of the harvest rate  $\lambda$, noise intensity $\epsilon$ and initial value $x_0$ on the stationary density function of the Fokker-Plank equation for the SDE in (\ref{log-SDE}). In section $\ref{Sec4}$, we review the deterministic logistic-harvest model with Allee effect (\ref{Allee-h}). We discuss the effect of the harvest rate $\lambda$, noise intensity $\epsilon$, and initial value $x_0$ on the stationary density function of the Fokker-Plank equation for the SDE in (\ref{log-SDE}). The Euler-Maruyama approximation is then  used to approximate the solution of the stochastic model. In  section $\ref{Sec5}$, we summarize numerical experiments to reveal the sample path behaviors of the deterministic and stochastic models. Finally, in section $\ref{Sec6}$, we present a short conclusion about our findings.
\section{Preliminaries}\label{Sec2a}
In this section, we recall some basic facts about Brownian motion and a stochastic differential equations.

Assume $(\Omega, \mathfrak{F}, \{\mathfrak{F_t}\}_{t>0}, \mathcal{P})$ is a complete probability space with a filtration $\{\mathfrak{F_t}\}_{t>0}$ satisfying the usual conditions, i.e. $\{\mathfrak{F_t}\}_{t>0}$ is increasing and continuous while $\mathfrak{F_0}$ contains all $\mathbb{P}-$null sets. Brownian motion $B_t$ is an abstract of random walk process \cite{Arnold(2003)} defined on the filtered probability space $(\Omega, \mathfrak{F}, \{\mathfrak{F_t}\}_{t>0}, \mathcal{P})$ which satisfies the following properties:
\begin{itemize}
\item  Stationary and normal increments: $B_t-B_s$, for $s<t$ is normally distributed with mean is equal to zero and variance is equal to $t - s,$
\item Independence of increments: $B_t-B_s$, for $s<t$, is independent of the past,
\item  Continuity of paths: $B_t$ is a continuous function of $t$, almost surely.
\item  The process starts at origin: $B_0 = 0$,
\item Brownian motion is nowhere differentiable, almost surely.
\end{itemize}
Stochastic differential equations \cite{Fima C2005introduction} are  often  used in modeling biological phenomena, by taking the intrinsic random effects into account. Intrinsic forcing induced SDE models are considered in population dynamics, epidemics, genetics, and oncogenesis.

Consider a stochastic differential equation driven by Gaussian noise
 \begin{equation}\label{SDE}
dX_t=f(X_t)dt+g(X_t)dB_t, \qquad t\in(0,\infty). \tag{2.1}
\end{equation}
If both drift  $f$ and noise intensity $g$ satisfy  a local Lipschitz condition, a growth condition or a priori  estimate on the solution, then stochastic differential equation (\ref{SDE}) has a unique continuous solution $X_t$ on $t\in(0,\infty).$ \cite{A. Friedman(1976),Applebaum Book(2009),duan2015introduction,L. Arnold(1972)}.
\section{\bf{Logistic-harvest model without Allee effect}}\label{Sec3}
\subsection{\bf{Deterministic logistic-harvest model without Allee effect }}\label{Sec31}
Consider a population $X_t$ with dynamic according to the logistic growth model without Allee effect. The idea is how to guarantee maximum stable yield in a resource population harvested at rate $\lambda X_t$ members per unit time. The harvested population in model (\ref{log-h}) can be written as:
\begin{equation}\label{LGH1}
\frac{dX_t}{dt} = r_1X_t\left(1-\frac{X_t}{K_1}\right),  \qquad X(0)=x_0, \tag{3.1}
\end{equation}
where $r_1=r-\lambda$ and $K_1=(1-\frac{\lambda}{r})K$. We define $F(x)=   r_1 x \left(1-\frac{x}{K_1}\right)$.

\begin{figure}[h!]
\begin{subfigure}[b]{0.5\linewidth}
  \includegraphics[width=\linewidth]{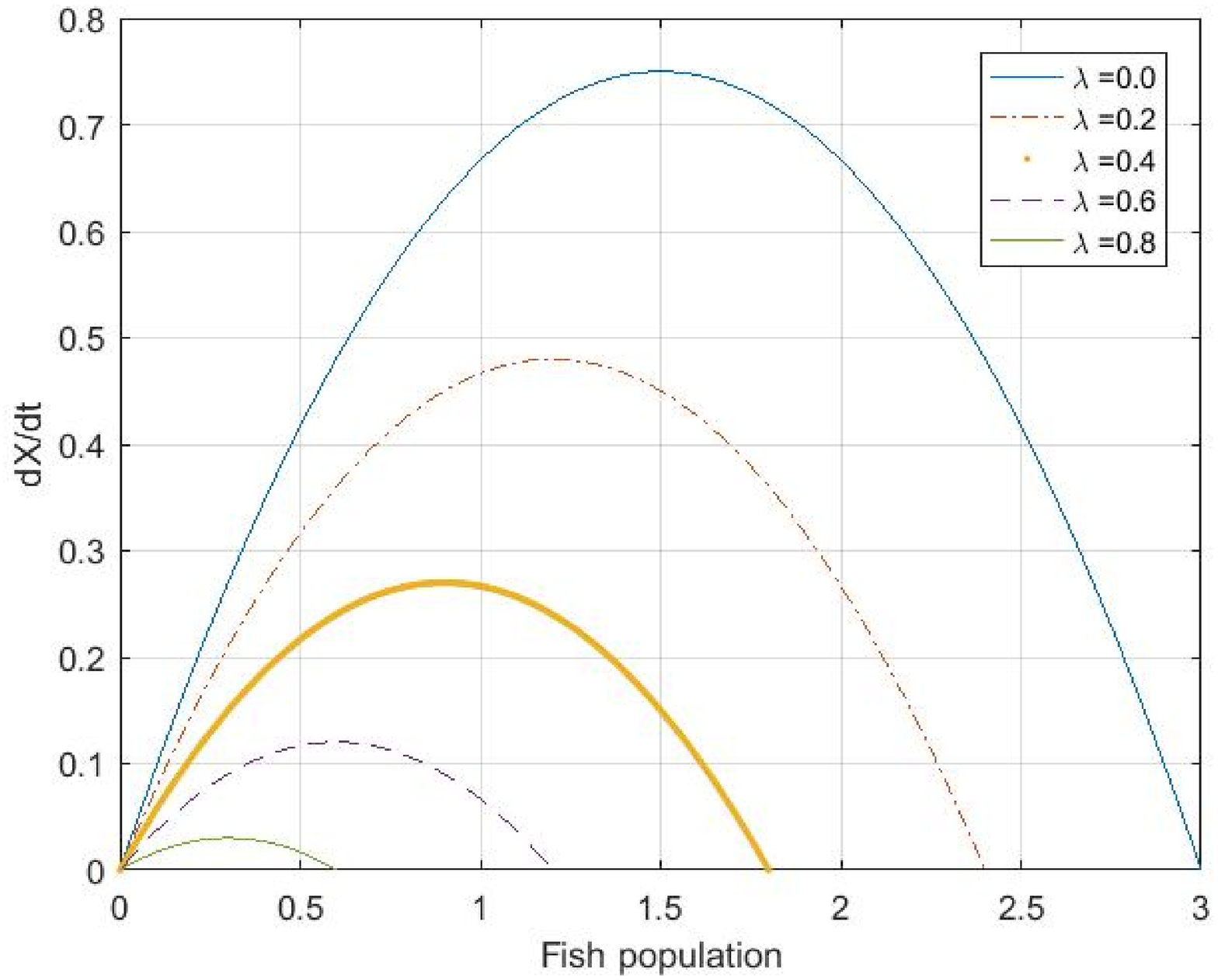}
 \caption{$\lambda$ vary.}
     \end{subfigure}
     \begin{subfigure}[b]{0.5\linewidth}
  \includegraphics[width=\linewidth]{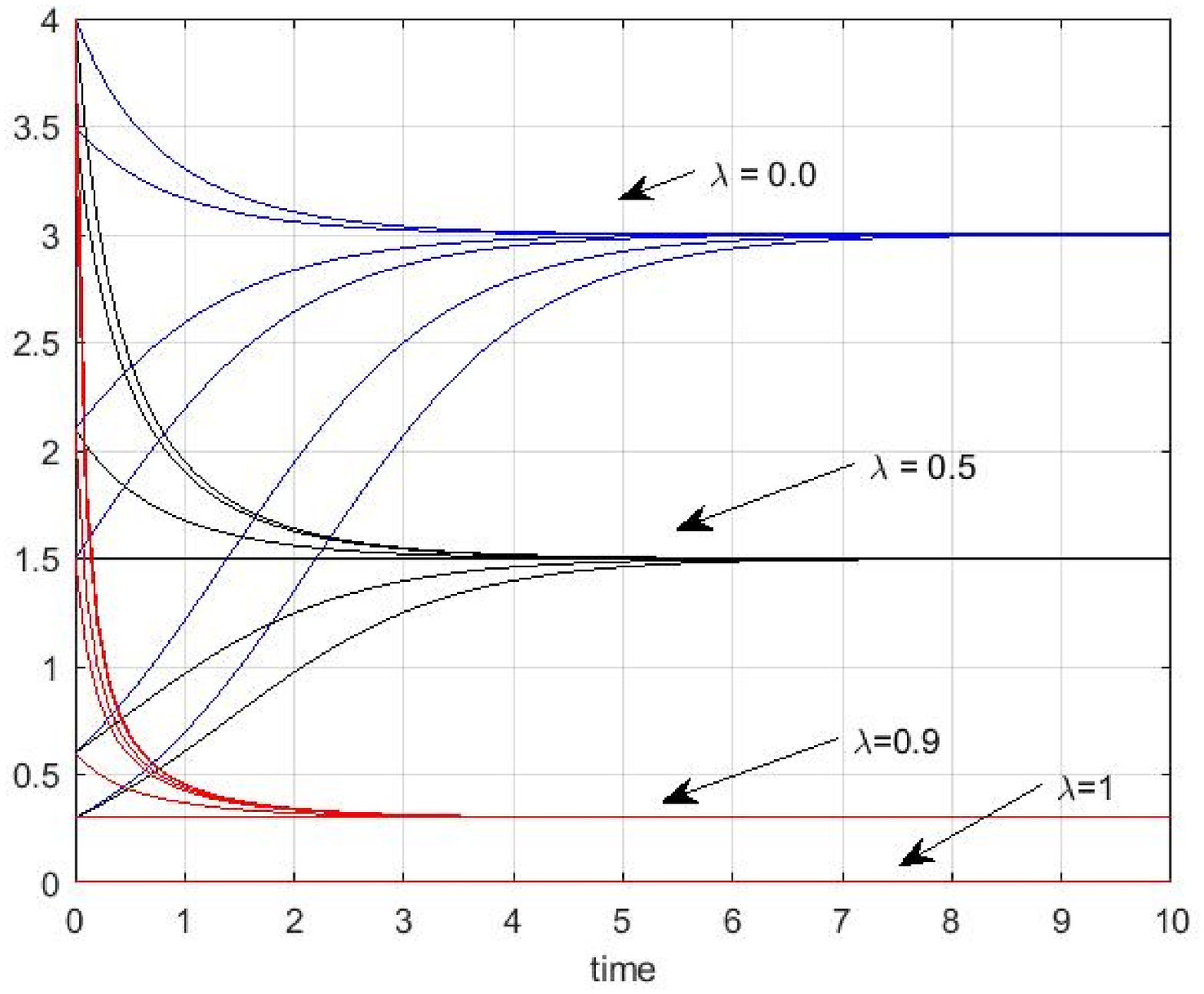}
 \caption{$\lambda$ and initial value vary.}
     \end{subfigure}

 \caption{The phase line and trajectories of $\frac{dX_t}{dt} = rX_t(1- X_t/K)-\lambda X_t$. (a) As the value of harvesting effort is sufficiently big ( $\lambda > r$), the population extinction occurs. (b) The solution of model (\ref{log-SDE}) for different value of $\lambda$ and $x_0$. Here we can see that as $\lambda$ increases, the population size $X_t$ goes to zero and $X_t$ has $S-$ shape when $x_0 < \frac{K_1}{2}.$ While $\frac{K_1}{2}< x_0 <K_1$ and $x_0>K_1$, the population size approaches to $K_1$ as $t\rightarrow\infty$.}
  \label{logD}
  \end{figure}

Equilibria points or constant solutions of (\ref{LGH1}), are $X_u = 0$  which is the trivial equilibrium point, and  $X_s= K( 1-\frac{\lambda}{r}),$ which is a non-trivial equilibrium point if $\lambda < r$. $X_u$ is unstable while $X_s$ is stable. For  $\lambda > r$, i.e. if the harvesting effort is very large, the population will die out. In this case $X_u=0$ is the only realistic steady state (equilibrium point) which is stable. The non-trivial equilibrium point $X_s$ is an asymptotic growth value of the harvest population model. Since $1 < K$ for $r > \lambda$, this implies that the asymptotic values of harvesting population lower than the non-harvesting population; (See Fig  \ref{logD}).

The function $F(x)$ in model (\ref{LGH1}) is autonomous function, because it is independent of $t$ and it is continuously differentiable ( class of $C^1$). Thus it has a unique solution and its non-trivial solution to the initial value problem is \cite{J.Murray(2002),Rahmani M.H.DOUST(2015)}
\begin{equation}\label{Sol1}
  X_t=\frac{X_{0}K _1}{X_{0}+(K_1-X_{0})e^{-r_1t}}, \qquad X(0)=x_{0},\tag{3.2}
 \end{equation}
The non-trivial solution (\ref{Sol1}) goes to the asymptotic value $K_1$ as time goes to infinity, i.e., $\lim_{t\rightarrow\infty}X_t=K_1$, for any $X_{0} > 0$. Hence $X_u=0$ is unstable because small perturbations increasing $X$ makes $dX_t/dt > 0,$ which further increases $X_t$ and the population rises towards $K_1$ which is asymptotically stable. When $x_0 > K_1$, $dX_t/dt < 0$ the population decline towards $K_1$. The function $F(x)$ has maximum value at $\frac{r_1K_1}{4}$ which is obtained by substituting $X=\frac{K_1}{2}$ in Eq. (\ref{LGH1}).
The deterministic model (\ref{LGH1}) can be written as
  \begin{equation*}
  \frac{dX_t}{dt}=-\frac{\partial V}{dx}
  \end{equation*}
where $V$ is the potential function defined by
  \begin{equation*}
  V(x)=-\int r_1x\left(1-\frac{x}{K_1}\right)dx= -\frac{-r_1}{2}x^2+\frac{r_1}{3K_1}x^3.
  \end{equation*}
The potential function has a local minimum corresponding to the stable equilibrium and a local maximum $x=0$ which is an unstable equilibrium if $r-\lambda >0$. The system has only one stable equilibrium, so it is called monostable.

For $r-\lambda >0$, the population converges to the stable equilibrium $X_s=K\left(1-\frac{\lambda}{r}\right)$, and the yield at the stable equilibrium, called the sustainable yield is $\bar K=\lambda X_s=\lambda K\left(1-\frac{\lambda}{r}\right).$ From this we can calculate the fishing effort that maximize is $\lambda_{MSY}=\frac{r}{2}$ which is called maximum sustainable yield (MSY) is $\bar \lambda =\frac{r_1K_1}{4}$, and the corresponding stable equilibrium is $K_{max}=\frac{K_1}{2}$.

From Figure \ref{logD}b, we can observe that when $\lambda=0$ and $x_0<\frac{K}{2}$, the phase point moves faster and faster until it reaches $\frac{K}{2}$, and $\frac{dX}{dt}$ reaches its maximum value $\frac{rK}{4}$. While the phase point approaches to wards carrying capacity $K$ if $\frac{K}{2}< x_0<K$ and $x_0>K$.

When $\lambda\neq 0$ and the initial value below half of the asymptotic value ($\frac{K_1}{2}$), the phase point moves faster and faster until it reaches $\frac{K_1}{2}$, and $\frac{dX}{dt}$ reaches its maximum value $\frac{r_1K_1}{4}$. While if $\frac{K_1}{2}< x_0<K_1$ and $x_0>K_1$, the phase point goes to wards $K_1$.

In a biological view, this tells us that the population initially growth faster and faster \cite{C.Braumann(2019)} and the graph of $X_t$ is concave up. But $\frac{dX}{dt}$ starts to decrease  if the initial value passes half of carrying capacity $K$ or half of asymptotic value $K_1$. In this case, $X_t$ has concave down shape. For initial value below half of carrying capacity $\frac{K}{2}$ or half of asymptotic value $\frac{K_1}{2}$, $X_t$ has $S$-shaped; ( see Figure \ref{logD}b ).
\subsection{\bf{Stochastic logistic-harvest model without Allee effect}}\label{Sec32}
We will consider stochastic perturbation of the logistic-harvest model without Allee effect (\ref{log-SDE}).
\begin{equation*}\label{log-SDE1}
dX_t = \left[rX_t\left(1-\frac{X_t}{K}\right)-\lambda X_t\right]dt+\epsilon X_tdB_t,  \qquad X_0=x_0.\tag{3.3}
\end{equation*}
Eq. (\ref{log-SDE1}) can be transformed into the form of the SDE as in our previous paper \cite{Almaz T.(2020)} and rewritten as
\begin{equation}\label{log-SDE1}
dX_t = (r-\lambda )X_t \left(1-\frac{X_t}{\left(1-\frac{\lambda }{r}\right)K}\right)dt+\epsilon X_tdB_t,  \qquad X_0=x_0.\tag{3.4}
\end{equation}
Since this model has four parameters, we non-dimensionalize by rescaling the population size (variable) and time. Then the new model (or SDE) will have fewer parameters. Because studying the qualitative behaviour of a SDE (or model) with many parameters is difficult. Define
\begin{equation*}
  Y=\frac{X_t}{K\left(1-\frac{\lambda}{r}\right)}=\frac{X_t}{K_1}, \quad \tau=(r-\lambda).t=r_1 t
\end{equation*}

The new model becomes
\begin{equation}\label{LGBM10}
dY = Y(1-Y)d\tau+\epsilon YdB\left(\frac{\tau}{r-\lambda}\right),  \qquad Y_0=\frac{x_0}{K\left(1-\frac{\lambda}{r}\right)}=y_0,\tag{3.5}
\end{equation}
or
\begin{equation}\label{LGBM1}
dY = Y(1-Y)d\tau + \frac{\epsilon}{\sqrt{r-\lambda}} YdB_\tau, \qquad Y_0=y_0,\tag{3.6}
\end{equation}
where $\epsilon$ is a positive constant representing random growth effects ( $0 <\epsilon <1$), and $r > \lambda$. $B_\tau$ is a Brownian motion which has independent and stationary increments with stochastically continuous sample paths.

The solution of the model \ref{LGBM1} is a homogenous diffusion process with the drift coefficient $\mu(t,y)=y(1-y)$ and diffusion term $\upsilon (t,y)=\frac{\epsilon^2}{r-\lambda} y^2$. Finding the exact solution of the nonlinear SDE in (\ref{LGBM1}) is similar with [\cite{V.M(2011)}, Section 9.3]. Set a new variable $Z=\frac{1}{Y}$ and apply It\^o formula \cite{E. Allen,V.M(2011)}. Our goal is that to reduce the nonlinear SDE in terms of $Y$ in to a linear SDE in $Z$, which we then able to solve. Thus we get a new linear SDE
\begin{equation}\label{z}
dZ=\left[\left({\frac{\epsilon^2}{{r-\lambda}}} - 1\right)Z +1\right]d\tau - \frac{\epsilon}{\sqrt{r-\lambda}} ZdB_\tau,\tag{3.7}
\end{equation}
 $$ Z_0=\frac{1}{n_0}.$$
According \cite{Simon(2019)} and [\cite{,{V.M(2011)}}, Theorem 9.4] , the solution of Eq. (\ref{z}) is
\begin{figure}[h!]
\begin{subfigure}[b]{0.5\linewidth}
  \includegraphics[width=\linewidth]{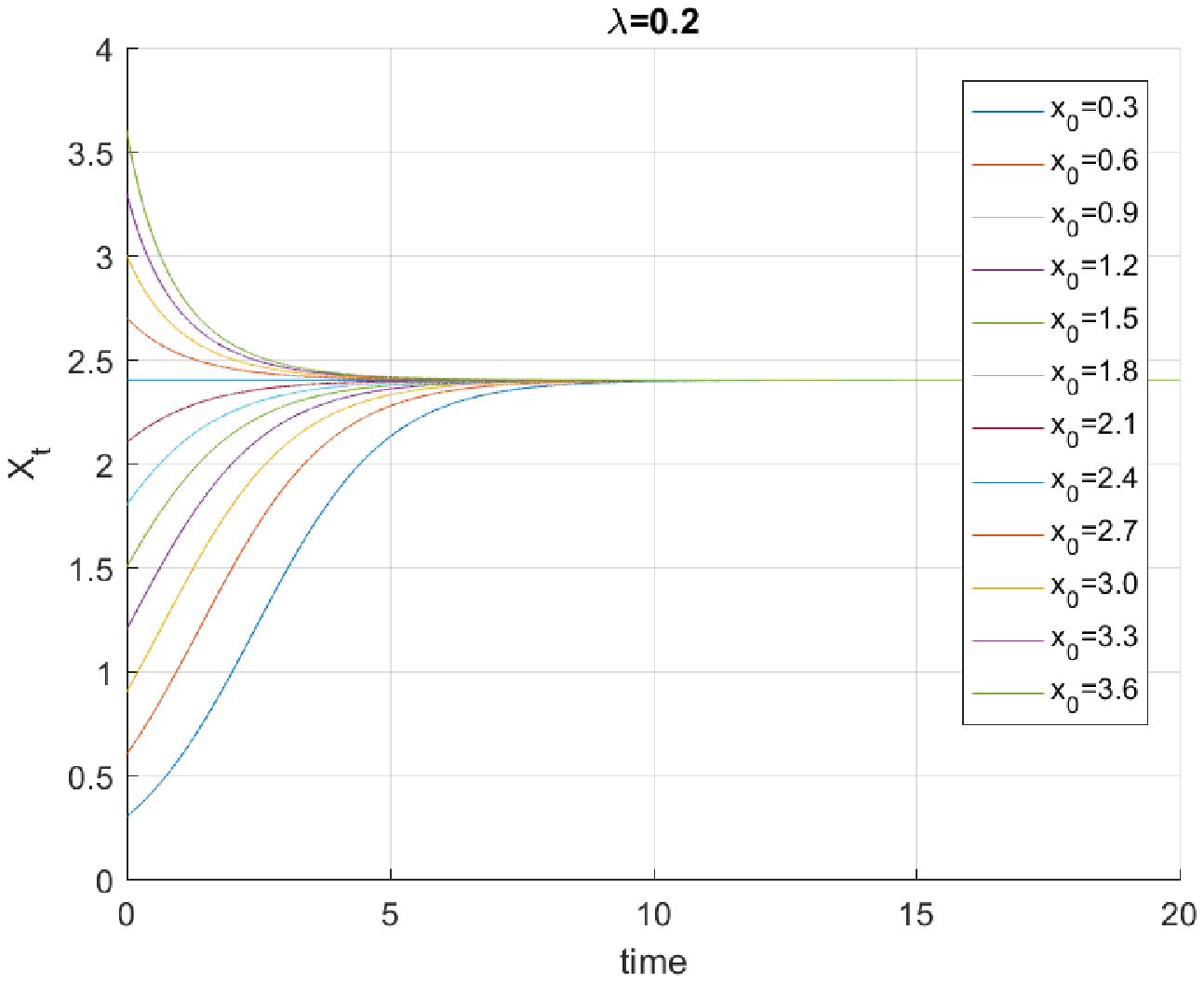}
 \caption{$\lambda = 0.2$ and $\epsilon=0.0$ and $x_0$ vary.}
     \end{subfigure}
    \begin{subfigure}[b]{0.5\linewidth}
  \includegraphics[width=\linewidth]{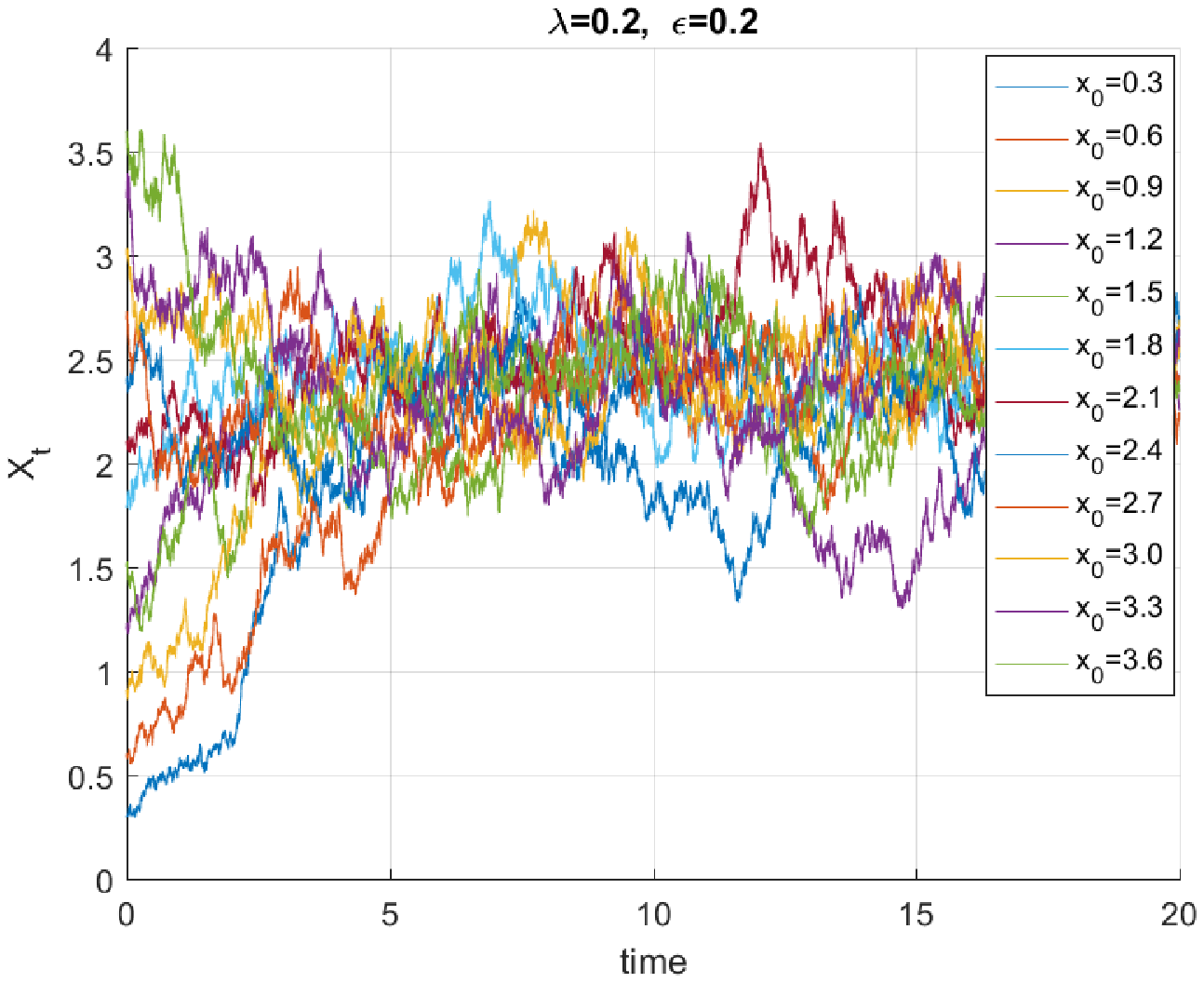}
 \caption{$\lambda = 0.2$, $\epsilon=0.2$  and $x_0$ vary.}
     \end{subfigure}
 \caption{Sample solutions of  $dX_t=\left[rX_t\left(1-\frac{X_t}{K}\right)-\lambda X_t\right]dt + \epsilon X_tdB_t$. (a) $\lambda = 0.2$, $\epsilon=0.0$ ( no noise) and initial value $x_0$ vary. (b)  $\lambda = 0.2$, $\epsilon=0.2$ and initial value $x_0$ vary. When $x_0 \in (0,K)$ and $x_0>K$, the population approaches its maximum population size. Parameters $r=1$, $K=3$. }
  \label{logsim}
\end{figure}
\begin{equation*}
Z=\varphi _{\tau}\left(Z_0+\int_{0}^{\tau}\varphi _{s}^{-1}ds\right), \qquad \tau \geq 0,
\end{equation*}
where $\varphi _{\tau}=\exp\left(\left(\frac{1}{2}\left(\frac{\epsilon}{\sqrt{r-\lambda}}\right)^2-1\right)\tau - \frac{\epsilon}{\sqrt{r-\lambda}}B_\tau\right).$

Since $Y=\frac{1}{Z}$,  we obtain the unique, strong solution of equation (\ref{LGBM1})
\begin{equation}\label{FinalSol}
Y=\frac{y_0\exp\left(\left(1-\frac{1}{2}\left(\frac{\epsilon}{\sqrt{r-\lambda}}\right)^2\right)\tau +\frac{\epsilon}{\sqrt{r-\lambda}} B_\tau\right)}{\left(1+y_0\int_{0}^{t}\varphi _{s}^{-1}ds\right)}. \tag{3.8}
 \end{equation}

From equation (\ref{FinalSol}), we observe that the solution exists for all $\tau>0$ and if $y_0 >0$, then $Y > 0$ a.s. If $\sqrt2 < \frac{\epsilon}{\sqrt{r-\lambda}}$, then $\left(1-\frac{1}{2}\left(\frac{\epsilon}{\sqrt{r-\lambda}}\right)^2\right)\tau +\frac{\epsilon}{\sqrt{r-\lambda}} B_\tau=\left[\left(1-\frac{1}{2}\left(\frac{\epsilon}{\sqrt{r-\lambda}}\right)^2\right) +\frac{\epsilon}{\sqrt{r-\lambda}} \frac{B_\tau}{\tau}\right]\tau$ goes to $-\infty$ as time $\tau \rightarrow\infty$. According to the strong law of large numbers, we apply $\frac{B_\tau}{\tau}=0$ as $\tau\rightarrow\infty$. From this we have $Y \rightarrow 0$ as $\tau\rightarrow\infty$.

When the value of Gaussian noise intensity $\epsilon$ is small, the solution in (\ref{FinalSol}) become a solution of the deterministic model in (\ref{log-h}), i. e., $$\lim_{\epsilon \rightarrow 0}Y=\frac{1}{1+\left(\frac{1}{y_0}-1\right)e^{-\tau}}.$$

The Euler-Maruyama method was implemented \cite{D. Heath} in order to give an approximation for the sample paths solution of the stochastic
model. Some sample solution paths are plotted in Figure  \ref{logsim}. We   observe that the sample  solution paths are positive.
\subsection{\bf{Extinction probability  }}\label{s32}
This subsection deals with the transition density function $p(y,\tau)$ for the process $Y=\{ Y_\tau, \tau > 0\}$ which satisfies the following theorem. The stationary density gives  important  long time information about the probabilistic behaviour of the solution of a given SDE.
\begin{theorem}\label{FPE}
( Fokker-Plank equation (FPE)): [ Simon (2019) \cite{Simon(2019)}, Theorem 5.4]. The probability density $p(x,t)$  of the solution of the SDE in (\ref{LGBM1}) solves the partial differential equation
\begin{equation}
\frac{\partial p}{\partial \tau}=- \frac{d}{dy}(y(1-y)p) +\frac{\gamma^2}{2} \frac{d^2}{dy^2}(y^2p),\tag{3.9}
\end{equation}
\end{theorem}
where $\gamma=\frac{\epsilon}{\sqrt{r-\lambda}}$ and with initial condition $p(y_s|y_\tau)=\delta (y_\tau-y_s)$ for $ \tau \geq s$.\\
\textbf{Proof}: See Simon (2019) \cite{Simon(2019)}.

\begin{figure}[h!]
\begin{subfigure}[b]{0.5\linewidth}
  \includegraphics[width=\linewidth]{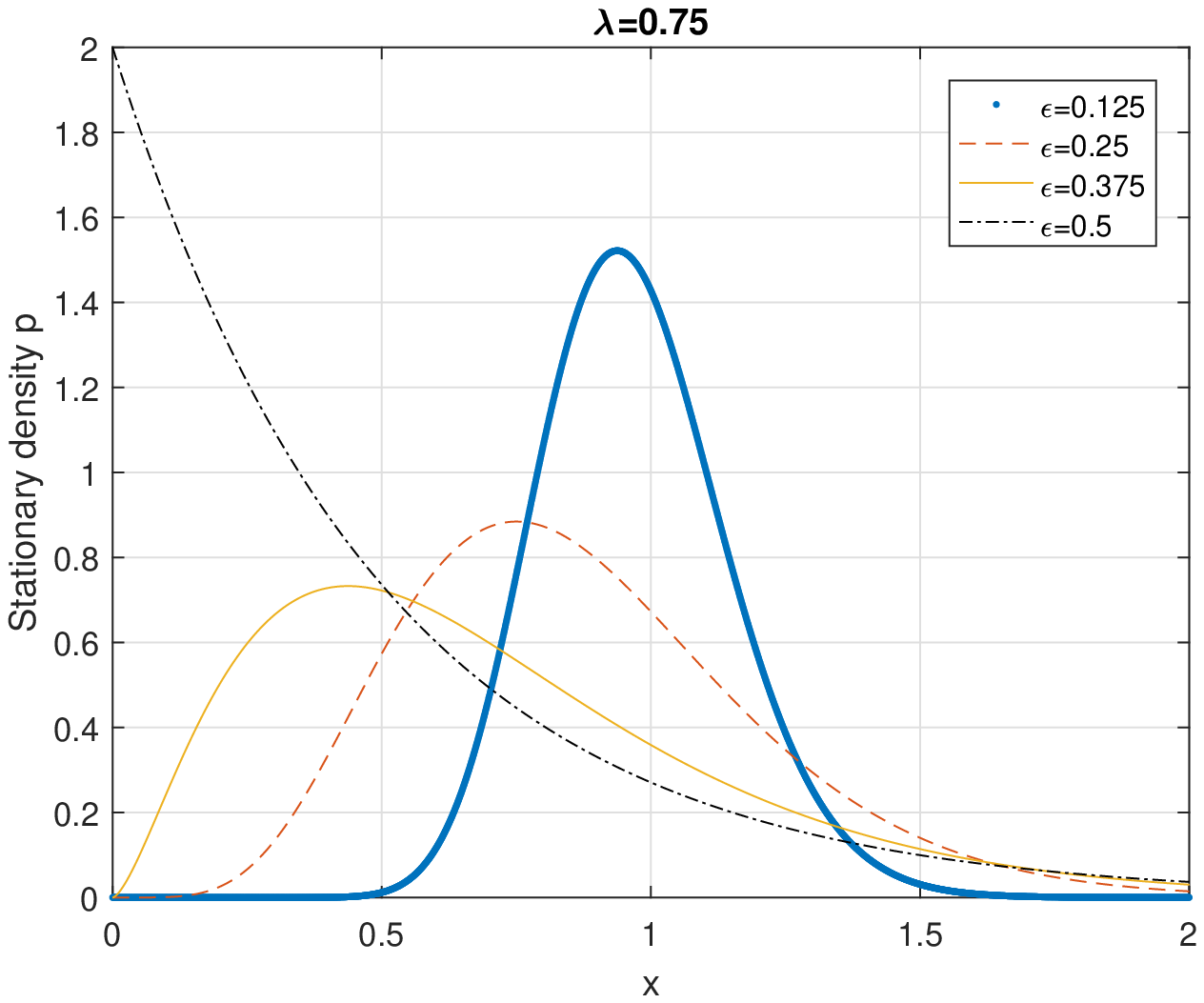}
 \caption{$\lambda = 0.75$ and $\epsilon$ vary.}
     \end{subfigure}
    \begin{subfigure}[b]{0.5\linewidth}
  \includegraphics[width=\linewidth]{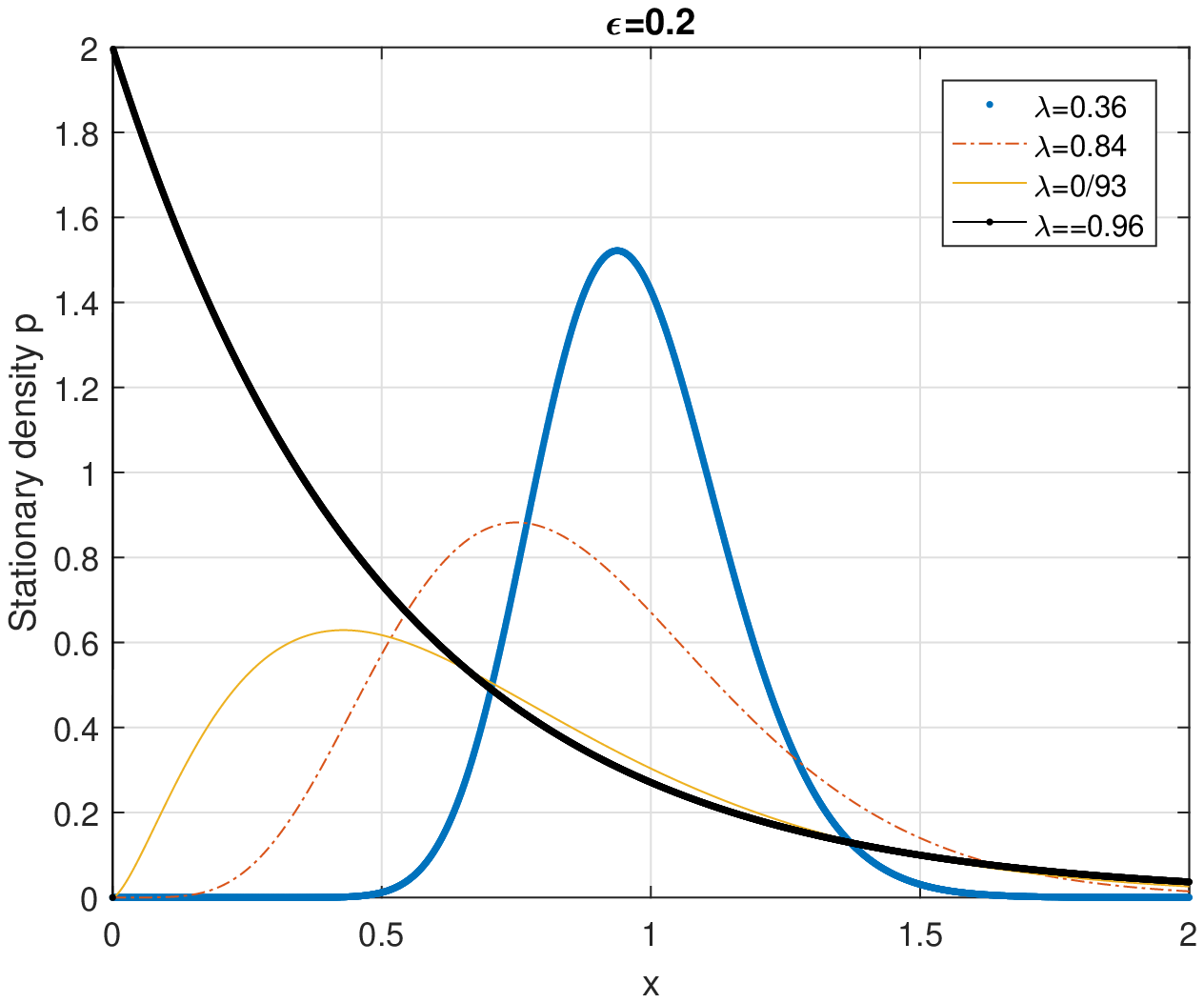}
 \caption{$\epsilon=0.2$ and $\lambda$ vary.}
     \end{subfigure}
  \caption{Stationary densities of model (\ref{log-SDE}) for $r=1$. (a) $\lambda=0.75$, $\epsilon=0.125, 0.25, 0.375, 0.5$. (b) $\epsilon=0.2$, $\lambda=0.36, 0.84, 0.93, 0.96$.}
  \label{pdf}
 \end{figure}
In our case, the  density $p$ satisfies the time-independent FPE. i.e. Eq. (\ref{FPE}) is the second order differential equation  as in  Mackeri$\breve{c}$ius \cite{V.M(2011)}
$$\frac{d}{dy}(y(1-y)p) - \frac{\gamma^2}{2}  \frac{d^2}{dy^2}(y^2p)=0.$$
Noting that $p(y)\geq 0$, for all $y\in(0,\infty)$, \cite{P.E.Kloeden} and $\int_{0}^{\infty}p(y)dy=1$. For $1 \geq \frac{\gamma^2}{2}$, the stationary density $p$ in $(0,\infty)$ is
\begin{equation*}
p(y)=M \; y^{\frac{2}{\gamma^2}-2}e^{\frac{-2y}{\gamma^2}}
\end{equation*}
(here $M$ is the normalizing constant). In $(0,\infty)$, the function $p$ is integrable if $\frac{2}{\gamma^2}-2 > -1$ or equivalently $\gamma^2 < 2$.
Setting $\lambda=0$ ( non-harvesting) recovers the logistic growth model which is widely studied in our first paper \cite{Almaz T.(2020)}. The authors derived the exact solution of SDE in (\ref{log-SDE}) for $\lambda=0$ and discussed about the qualitative behaviour of the solution of the Fokker-Plank equation. This present work focuses for $\lambda\neq0$ ( harvesting case).

When $\gamma^2 \geq 2$ the diffusion process (SDE) in (\ref{LGBM1}) has no stationary density. That means population becomes extinct, but we have a noise-induced transition for $0<\gamma<\sqrt2$. In this case extinction can not occurs; ( Figure \ref{pdf}). In fact
\[
\lim_{y\rightarrow 0} p(y)=
\begin{cases}
0, \qquad  \qquad 1>\gamma^2\\
M=2, \qquad  1=\gamma^2\\
\infty, \qquad \qquad  \frac{\gamma^2}{2}< 1<\gamma^2\\
\end{cases}
\]

The next step is to show how to find the maximum point $y_{max}$ of $p(y)$. Using $p'(y)=0$ \cite{V.M(2011)}we can easily find $y_{max}$, so we have $y_{max}=1-\gamma^2$. When $\gamma$ becomes small, $y_{max}=1$. In this case the stationary density become more acute.
\section{\bf{Logistic-harvest model with Allee effect}}\label{Sec4}
\begin{figure}[h!]
\begin{subfigure}[b]{0.5\linewidth}
  \includegraphics[width=\linewidth]{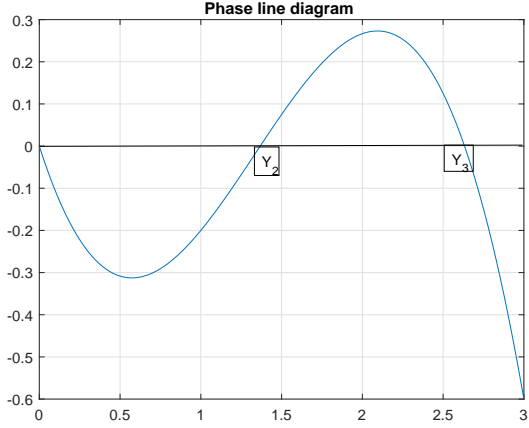}
 \caption{ $ \lambda = 0.2$.}
     \end{subfigure}
     \begin{subfigure}[b]{0.5\linewidth}
  \includegraphics[width=\linewidth]{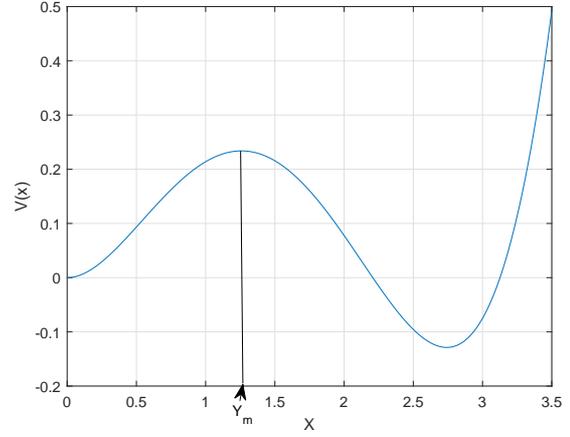}
 \caption{ $ \lambda = 0.15$.}
    \end{subfigure}
  \begin{subfigure}[b]{0.5\linewidth}
  \includegraphics[width=\linewidth]{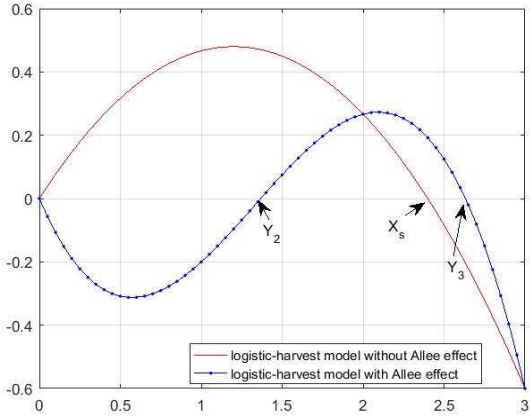}
 \caption{$\lambda = 0.2$.}
     \end{subfigure}
 \begin{subfigure}[b]{0.5\linewidth}
  \includegraphics[width=\linewidth]{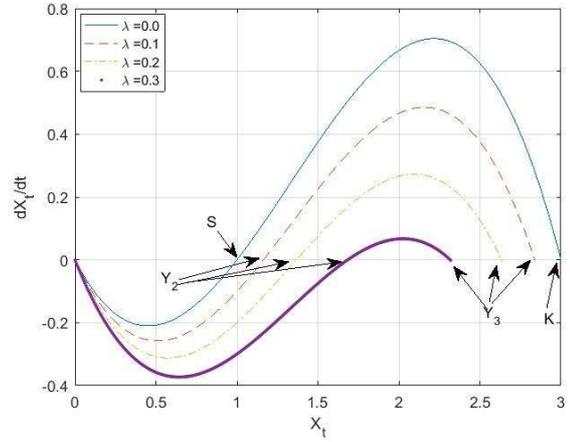}
 \caption{$\lambda$ vary.}
     \end{subfigure}
 \caption{ (a) Phase line diagram of (\ref{Allee-h}). (b)  Potential function $V$ of the model (\ref{Allee-h}). (c)  $\lambda=0.2$. In this cases, model (\ref{log-h}) is always positive while model (\ref{Allee-h}) is negative when the population $X_t < S.$  (d) $\lambda$ vary. Parameters $r=1$, $S=1$, $K=3$ }
  \label{logDet}
    \end{figure}
\subsection{\bf{Deterministic logistic-harvest model with Allee effect}}\label{Sec41}
 Now let's nondimensionalize the Allee effect model (\ref{Allee-h})
 \begin{equation*}
   \frac{dX_t}{dt}=rX_t\left(\frac{X_t}{S}-1\right)\left(1-\frac{X_t}{K}\right)-\lambda X_t,
 \end{equation*}
which helps to rescale variables such that the rescaled model has fewer parameters. Let's rescale population size $X_t$ by expressing it relative to the carrying capacity $K$ (scaling by $S$ would work as well).

Setting $Y_t=\frac{X_t}{K}$ and $\beta=\frac{K}{S}$. The new differential equation has the following form:
\begin{equation}\label{modif}
\frac{dY_t}{dt}=rY_t(\beta Y_t-1)(1-Y_t)-\lambda  Y_t, \qquad Y(0)=y_0,\tag{4.1}
\end{equation}
where $y_0=\frac{x_0}{K}$. Our new model has just three parameter, which makes the bifurcation analyses, computation of equilibria, etc. more transparent.

It is clear that if $\lambda  = 0$, then the logistic-harvesting  model with Allee effect in Eq. (\ref{Allee-h}) reduces to the non-harvesting logistic growth model with Allee effect as given in Eq. (\ref{Allee}).

Equation (\ref{modif}) has one trivial equilibria point at $Y_t = 0$  and two non-trivial equilibrium points at $ Y_t = \frac{\beta +1 \pm \sqrt{(\beta + 1)^2 - 4\beta (1+ \lambda /r)}}{2\beta}$  if $(\beta + 1)^2 - 4\beta (1+ \lambda /r) > 0$. Thus model (\ref{modif}) has the following three equilibrium points.\\
\begin{equation}\label{equi}
  Y_1=0, \quad Y_2=\frac{(\beta +1) - \sqrt{(\beta + 1)^2 - 4\beta (1+ \lambda /r)}}{2\beta}, \quad Y_3=\frac{(\beta +1) + \sqrt{(\beta + 1)^2 - 4\beta (1+ \lambda /r)}}{2\beta}.\tag{4.2}
\end{equation}
$Y_1$ and $Y_3$ are stable equilibria separated by unstable equilibrium $Y_2$. Set $m_1 = r(\frac{(\beta - 1)^2}{4\beta})$ which is called  the critical point.

Clearly, if  we use $\lambda > m_1$ in Eq. (\ref{equi}), no fixed point which shows $Y_1=0$ is the only equilibrium point which is stable. While if $\lambda  = m_1$, there exists two equilibrium points, i.e. $Y_1=0$ (stable) and $Y_m=\frac{\beta +1}{2\beta}$ (unstable); ( See Fig. \ref{logDet}c).

The non-trivial  equilibrium point $Y_3$ is an asymptotic growth value of the harvest model. Since $Y_3 < K$ for $\lambda  < m_1$, this implies that the asymptotic values of harvesting fish population lower than the non-harvesting fish population; ( See Fig. \ref{logDet}d ).

In Figure \ref{logDet}b, we plot the graph of the potential function $V(x)$ for values of $\lambda=0.15$, defined by
  \begin{equation*}
  V(x)=-\int \left[rx\left(\frac{x}{S}-1\right)\left(1-\frac{x}{K}\right)-\lambda x\right]dx.
  \end{equation*}

In term of $V(x)$, Eq. (\ref{Allee-h}) can be written as:
\begin{equation*}
\frac{dx}{dt}=-\frac{d V}{dx}.
\end{equation*}

The phase line diagram for Eq. (\ref{Allee-h}) is shown in Figure \ref{logDet}a. Denoting the stable equilibrium point by $Y_3$ and the unstable equilibrium point by $Y_2$, the separation between the two equilibrium points is $Y_3 -Y_2$.

If $\lambda < m_1$, Figure \ref{logDet}b shows the potential function $V(x)$ has two local minima corresponding to the stable equilibrium $Y_1$ and $Y_3$ and one local maximum at $Y_2$ which is an unstable  equilibrium.  The function $V(x)$ is called a double-well potential, because the two stable equilibrium $Y_1$ and $Y_3$ separated by an unstable  equilibrium $Y_2$.

From the biological point of view, it is meaningful to choose $\beta > 1$, and $0 < \lambda < r\left(\frac{(\beta + 1)^2}{4\beta} - 1\right)$ ( or $0 < \lambda  < m_1)$, and the state $Y_1$ represents to the population free state that means it is the state of population extinction, in this case no population are present. The state $Y_3$ implies the state of stable population, where the population density does not increase but stays at a constant level.

 The number of equilibrium points depends on the sign of $m$, where $m = (\beta + 1)^2 - 4\beta (1+ \lambda /r)$.\\
$(i)$ If $m > 0$, then there are two fixed points; (Two non-trivial equilibrium points),\\
$(ii)$ If $m = 0$, then there is only one point; (One  non-trivial equilibrium points),\\
$(iii)$ If $m < 0$, then there is no fixed point; (No non-trivial equilibrium points).\\

When $m < 0,$ i.e. $\lambda > m_1$ the population will go extinct as $t \rightarrow \infty$.  As far as biological meaning is concerned, we must catch at a harvest rate $\lambda < r\left(\frac{(\beta + 1)^2}{4\beta}-1\right)$. So in this case the model in (\ref{modif}) has two equilibria, one stable $Y_3$ and one unstable $Y_2$ with $Y_2 < Y_3$.
\begin{figure}[h!]
\begin{center}
  \includegraphics[width=0.5\linewidth]{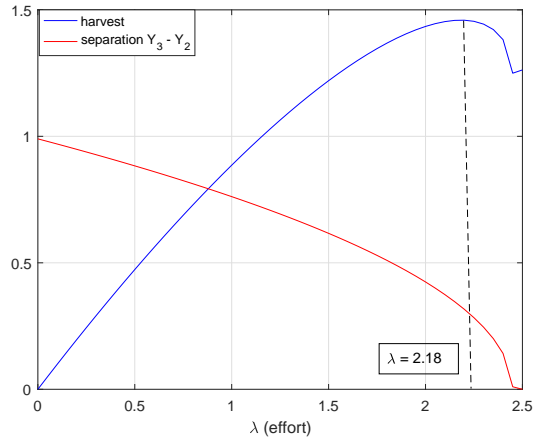}
 \caption{Blue: Harvest yield $\lambda Y_3(\lambda)$  versus  $\lambda$. Red: Separation $Y_3(\lambda)-Y_2(\lambda)$ versus $\lambda)$. }
 \label{max}
 \end{center}
 \end{figure}
In Figure \ref{logDet}c shows the phase line plots for Eq. (\ref{modif}), $\frac{dY_t}{dt}=rY_t(\beta Y_t-1)(1-Y_t)-\lambda  Y_t$ for increasing $\lambda$. If $\lambda$ is less than the critical point $m_1$, there are two stable equilibrium solutions and one unstable equilibrium solution. As $\lambda$ increases beyond $m_1$, there is one stable equilibrium solution.

Using $r = 0.1$ and $\beta = 100$, Figure \ref{max} shows plots of harvest yield $\lambda Y_3(\lambda)$  versus eﬀort $\lambda$ and separation $Y_3(\lambda)-Y_2(\lambda)$ versus $\lambda$.
Note that from Figure \ref{max} we get maximum yield when $\lambda = 2.18$. For the value of $\lambda$, the separation between the two equilibrium solutions is $Y_3(\lambda) -Y_2(\lambda) = 0.3283$. If there is noise in the system, we have to think that the expected time to extinction is not very long. This is a problem because while we want to maximize harvest yield, we do not want the stable and unstable equilibrium points to be close together because the expected time to extinction may be too short. Harvesting to maximize yield while driving the population to extinction is not a good harvesting strategy. It would be interesting to think about rational harvesting strategies that do not put the population in danger of extinction.
\subsection{\bf{Stochastic logistic-harvest model with Allee effect}}\label{Sec42}
We consider the dimensionaless stochastic perturbation of the logistic-harvest model with Allee effect (\ref{Allee-SDE}) by setting a new variable $Y_t=\frac{X_t}{K}$.
\begin{equation}\label{LGBM2}
dY_t = [rY_t(\beta Y_t-1)(1-Y_t)-\lambda Y_t]dt + \epsilon Y_tdB_t, \qquad Y(0)=\frac{x_0}{K}=y_0,\tag{4.3}
\end{equation}
where $\beta = \frac{K}{S} > 1, K$ is the carrying capacity and $S$ is the Allee parameter with $0 < S <K$. $B_t$ is a Brownian motion with stochastically continuous sample paths, as well as independent and stationary increments. The stochastic perturbation of the logistic-harvest model with Allee effort is discussed in \cite{A. Friedman(1976),Applebaum Book(2009),duan2015introduction,L. Arnold(1972)}.

\begin{figure}[h!]
 \begin{subfigure}[b]{0.5\linewidth}
  \includegraphics[width=\linewidth]{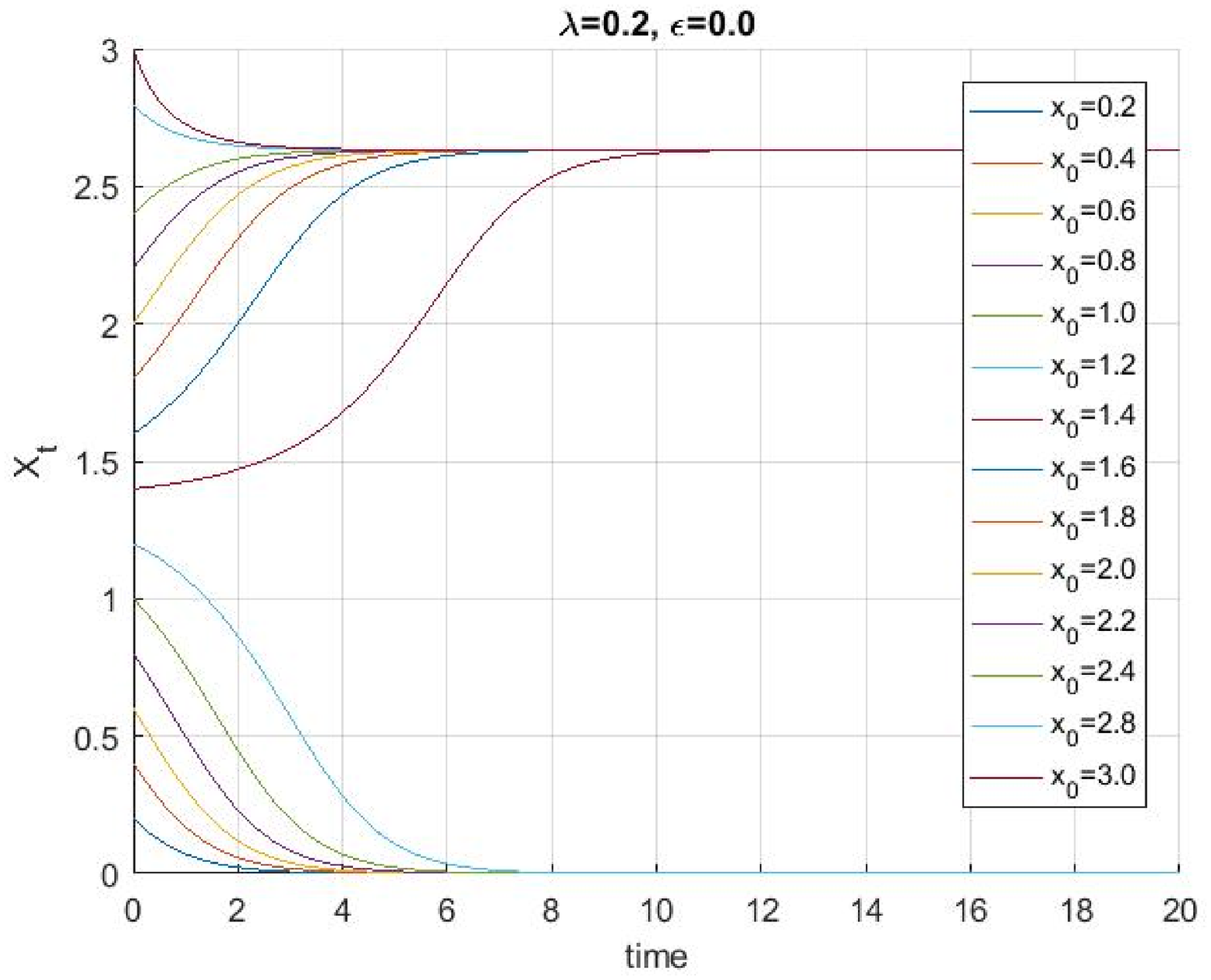}
 \caption{$\lambda=0.2$, $\epsilon=0$, and initial value $x_0$ vary.}
     \end{subfigure}
     \begin{subfigure}[b]{0.5\linewidth}
  \includegraphics[width=\linewidth]{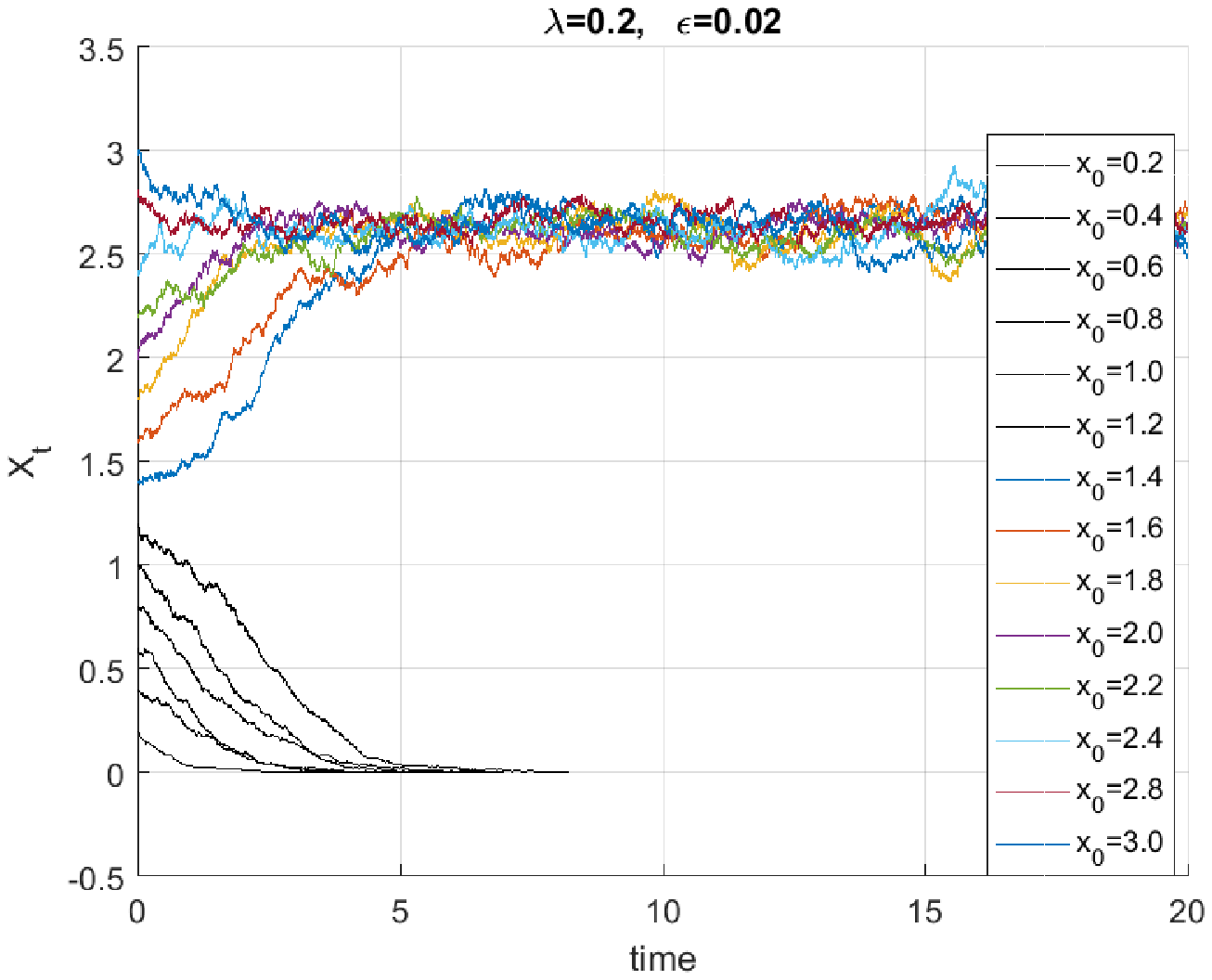}
 \caption{$\lambda=0.2$, $\epsilon=0.02$ and initial value $x_0$ vary.}
     \end{subfigure}
 \caption{Sample solutions of stochastic logistic-harvest model with Allee effect. (a) the solution of (\ref{LGBM2}) with $\lambda=0.2$, $\epsilon=0$ ( no noise) and initial value $x_0$ vary. (b) $\lambda = 0.2 $ and $\epsilon=0.02$ and $x_0$ vary. Here it is clearly seen that when $x_0 < S$, the population extinct occurs. While for $x_0 > S$ and $x_0 >K$, the population size approaches to its maximum size $K$. Parameters $r=1$, $S=1$, $K=3$.}
  \label{simulation}
\end{figure}
The non-trivial solution of the SDE in (\ref{LGBM2} can be found as follows \cite{Qingshan Yang(2011),M. Krstic(2010),Daqing Jiang(2005)}. Having in mind that $S < Y(0) < K$, let's define $C^2-$function $Z_t:R_{+}\rightarrow R_{+}$ as $Z_t=\log(Y_t)$, and apply It\^o formula to $Z_t$,  the system in (\ref{LGBM2}) is converted to a SDE with additive noise (to remove any state or level-dependent noise from these trajectories):
\begin{equation}\label{new}
dZ_t=f_1(Z_t)dt + g_1(Z_t)dB(t), \qquad t\in (0,T),\tag{4.4}
\end{equation}
where $f_1(Z_t)=r(\beta e^{Z_t}-1)(1-e^{Z_t})-\lambda -\frac{\epsilon^2}{2}$ , \qquad $g_1(Z_t)=  \epsilon.$

Note that  $f_1(Z_t)=0$ is the same as $f(Y_t)-\frac{\epsilon^2}{2}=0$. This shows the equilibrium point of the deterministic term of the additive noise system in (\ref{new})  is affected by the Gaussian noise intensity $\epsilon$.

Now we will show that  $Y_t$ is the solution of the SDE in (\ref{LGBM2}). Since $Y_t=e^{Z_t}$, apply It\^o formula to have
$$dY_t=de^{Z_t}=e^{Z_t}dZ_t + \frac{1}{2}e^{Z_t}(d{Z_t})^2$$
  \qquad  $$=e^{Z_t}\left[\left(r(\beta e^{Z_t}-1)(1-e^{Z_t})-\lambda -\frac{\epsilon^2}{2}\right)dt - \epsilon dB(t)\right] + \frac{1}{2}e^{Z_t}\epsilon^2 dt$$
\qquad  $$=Y_t[r(\beta Y_t-1)(1-Y_t)-\lambda]dt +\epsilon dB(t)].$$
This solution is strong, continuous and positive,     for $ S <Y(0)$ and $0< S < K.$

The numerical simulation (solution paths) of the stochastic differential equation  in  (\ref{LGBM2}) is  shown  in Figure \ref{simulation} with various initial values. To plot this we use the Euler-Maruyama method. As we can see in Figure \ref{simulation}, the sample path are positive and approaching to the carrying capacity $Y_3$ when $0< \lambda < \frac{1}{3}$. While it goes to extinction when $\lambda \geq \frac{1}{3}$. From  Figure \ref{simulation}c we observe that all trajectories, except $x=Y_2$ ( unstable equilibrium point) fall in to a potential pit ( $x=0$ and $x=Y_3$) the stable equilibrium points.

The Euler-Maryuama approximation was used to approximate the solution of stochastic model. Different values of the constant in the drift coefficient $\lambda$ were applied.

Next we prove that sample paths of $X_t$ of SDE (\ref{Allee-SDE}) are uniformly continuous for a.e. $t \geq 0$.  To show this consider the following integral
\begin{equation}\label{Int}
X_t = X(0)+\int_{0}^{t}f(X(s))dt + \int_{0}^{t}g(X(s))dB_s,\tag{4.5}
\end{equation}
where $f(X_s)=r X_s\left(\frac{X_s}{S}-1\right)\left(1-\frac{X_s}{K}\right)-\lambda X_s$,  $g(X_s)=\epsilon X_s$ and $0< S < X(0) < K.$

Suppose $0 < a < b < \infty$, $b-a \leq 1$, and $p > 2$. By applying the well known H\"{o}lder inequality and moment inequality for It\^o integrals (\ref{Int}), we have
\begin{equation}\label{LGBME}
\mathbf{E}|X_t-X_s|^{p}\leq2^{p-1}(b-a)^{p-1}\int_{a}^{b}\mathbf{E}[f(X_s)]^{p}ds+2^{p-1}\left(\frac{p(p-1)}{2}\right)^{\frac{p}{2}}\int_{a}^{b}\mathbf{E}[g(X_s]^{p}ds.\tag{4.6}
\end{equation}
Since,
$$\mathbf{E}[f(X_s)]^{p}\leq \left(\frac{K(K-S)}{S}\,r\right)^p+(\lambda K)^p,$$
$$\mathbf{E}[g(X_s)]^{p}\leq (K \epsilon)^p,$$
The equation in (\ref{LGBME}) can be estimated by
$$\mathbf{E}|X_t-X_s|^{p}\leq Q(b-a)^{\frac{p}{2}},$$
where $Q=2^{p-1}K^p((\frac{(K-S)}{S})^pr^p + \lambda ^p +(\frac{p(p-1)}{2})^{\frac{p}{2}}\epsilon^{p}).$

According to Kolmogorov-Centsov theorem on the continuity of a stochastic process \cite{M. Krstic(2010)},  we know  that almost every sample   path of $X_t$ is locally but uniformly H\"{o}lder-continuous with exponent $0<\gamma<\frac{p-2}{2p}$. Therefore the SDE in (\ref{Allee-SDE}) has uniformly continuous solution on $t\geq0$. All solutions of this model goes to zero as $t\rightarrow\infty$. Since $Y_t=\frac{X_t}{K}$, so the SDE in  (\ref{LGBM2}) has uniformly continuous solution on $t\geq0$ and its solution also approaches to zero as $t\rightarrow\infty$.
\subsection{\bf{Extinction probability and first passage probability}}\label{s42}
This subsection explains where the probability that the population extinct will happens, and the probability of reaching a large population size $L$ before reaching a small one.
Trajectories that start in the potential well on the right will eventually jump into the potential well on the left, even though it may take a very long time. Once there, they rapidly move to the region around $x = 0$ near the bottom of that well. Once there, they exit at zero with probability one. To see this, for small values of $Y_t$, Eq. (\ref{LGBM2}) can be approximated by
\begin{equation}\label{2}
dY_t=-(r+\lambda)dt+\epsilon Y_tdB_t.\tag{4.7}
\end{equation}

Then the boundary value problem for the probability $P(y)$ of exit at 0 before exit at $L$ is
\begin{equation}\label{prt}
  \frac{1}{2}\epsilon^2y^2p''-(r+\lambda)yp'=0, \qquad p(0)=1, p(L)=0, \tag{4.8}
\end{equation}
which has the solution
$$p(y)=1-\frac{y^{1+(r+\lambda)/\epsilon^2}}{L^{1+(r+\lambda)/\epsilon^2}}$$
Note that $P(y) \rightarrow 1$ as $L\rightarrow\infty$, even though we are using small $y$ approximations for values of $y$ that are not small. We would obtain the same result even if we solved the probability of exit problem corresponding to Eq. (\ref{LGBM2}).

According  to Theorem  \ref{FPE},  the Fokker-Planck equation corresponding to Eq. (\ref{LGBM2}) is
\begin{equation}\label{fpkAlle}
  \frac{\partial p}{\partial t}=\frac{1}{2}\epsilon^2 \frac{\partial^2}{\partial y^2}[y^2p]-\frac{{\partial }}{\partial y}[ry(\beta y-1)(1-y)-\lambda y)p]. \tag{4.9}
\end{equation}
Since all solutions of Eq. (\ref{LGBM2}) go to zero with probability one no matter what the starting point $y_0$, we expect that the solution of Eq. (\ref{fpkAlle}) satisfies
$$\lim_{t\rightarrow\infty} p(y,t/y_0), \qquad 0< y_0<\infty.$$
\begin{figure}[h!]
\begin{center}
  \includegraphics[width=0.6\linewidth]{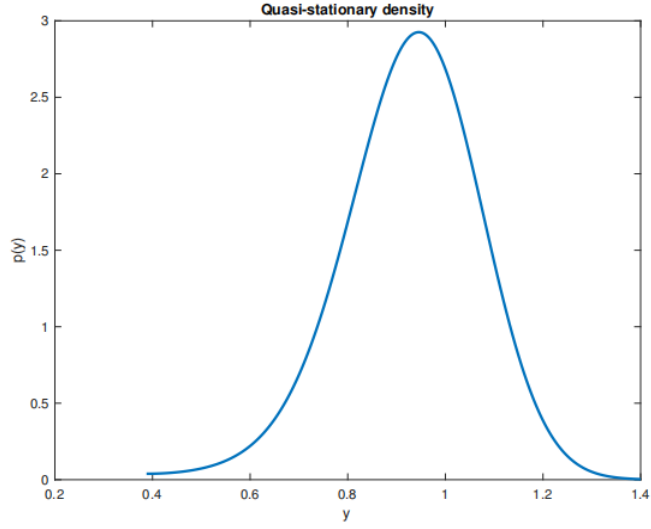}
 \caption{ The quasi-stationary density of model (\ref{LGBM2})}
 \label{qr}
 \end{center}
\end{figure}
In other words, all populations eventually become extinct. However, it is reasonable, or realistic, values of the parameters, if $y_0$ is in the potential well on the right in Figure \ref{logDet}b, it will take a very long time before the trajectory jumps across the potential barrier into the potential well on the left. In this case it makes sense to look at a quasi-stationary density \cite{V.M(2011)}, say $q(y)$, that is obtained by solving
\begin{equation*}
 \frac{1}{2}\epsilon^2 \frac{\partial^2}{\partial y^2}[y^2p]-\frac{{\partial }}{\partial y}[ry(\beta y-1)(1-y)-\lambda y)p]=0
\end{equation*}
on the interval $[Y_m, \infty]$ where $Y_m$ is the location of the local maximum of the potential function, shown at the vertical blank line in Figure \ref{logDet}b. The quasi-stationary density $q(y)$, given by

$$q(y)=\frac{y^{-2-2(r+\lambda)/\epsilon^2}e^{[-r\beta y^2+2r(1+\beta)y]/\epsilon^2}}{\int_{Y_m}^{\infty}y^{-2-2(r+\lambda)/\epsilon^2}e^{[-r\beta y^2+2r(1+\beta)y]/\epsilon^2}dy}   $$
is shown in Figure \ref{qr}.
\section{Numerical experiments}\label{Sec5}
We summarize our numerical findings  about  the impact of parameters $x_0$, $\lambda$ and $\epsilon$ on the solution of the deterministic and stochastic models of logistic-harvest with and without Allee effect.

Here we apply the Euler-Maruyama (EM) method following \cite{D. Higham (2001)} to Eq. (\ref{LGBM1}). To apply this method in the SDE (\ref{LGBM1}) over time $[0, T]$, we first need discretize the interval. For any positive $n$ assume $\Delta t= T/n$, and $s_j=j t$, for $j=1,2,...,n$. The numerical approximation to the solution $X(s_j)$ is denoted by $X_j$. As in \cite{D. Higham (2001)}, the EM method has the following form:
\begin{equation}\label{EM}
X_j=X_{j-1}+f(X_j)\Delta t+g(X_j)(B(X_j)-B(X_{j-1}), \qquad  j=1,2,3,.., n. \tag{5.1}
\end{equation}
\subsection{ \textbf{Numerical results and biological implications of logistic-harvest model without Allee effect}}
The phase line and trajectories of $\frac{dX_t}{dt} = rX_t(1- X_t/K)-\lambda X_t$ is plotted in Figure \ref{logD}. Parameters $r=1$, $K=3$ and $0 \leq\lambda \leq r.$  In Fig. \ref{logD}a as the value of harvesting effort is sufficiently large ( overfishing ), the population extinction occurs. Here the value of $X_u$ becomes small as $\lambda$ increases. In Fig. \ref{logD}b the solution of model (\ref{log-SDE}) for different values of $\lambda$ and $x_0$. In this figure, we can observe that as $\lambda$ increases, the population size $X_t$ decreases i.e. $X_t$ goes to zero and it has $S-$shape when $x_0 < \frac{K_1}{2}$. While $\frac{K_1}{2}< x_0 <K_1$ and $x_0>K_1$, the population size approaches to $K_1$ as $t\rightarrow\infty$. When $\lambda\neq 0$ and $x_0<\frac{K_1}{2}$, the phase point moves faster and faster until it reaches $\frac{K_1}{2}$, and $\frac{dX}{dt}$ reaches its maximum value $\frac{r_1K_1}{4}$. While if $\frac{K_1}{2}< x_0<K_1$ and $x_0>K_1$, the phase point goes to wards $K_1$. The biological implications of this result tells us that the population initially grows faster and faster \cite{C.Braumann(2019)} and the graph of $X_t$ is concave up. But $\frac{dX}{dt}$ starts to decrease  if $x_0>\frac{K}{2}$ or $x_0>\frac{K}{2}$. In this case, the shape of $X_t$ is concave down.

Figure \ref{logsim} shows the numerical simulation of model $dX_t=[rX_t(1-\frac{X_t}{K})-\lambda X_t]dt + \epsilon X_tdB_t$ with fixed parameters $r=1$, $K=3$. For $\lambda = 0.2$, $\epsilon=0.0$ ( no noise) and initial value $x_0$ vary is plotted in Fig \ref{logsim}a. Fig. \ref{logsim}b presents numerical simulation of stochastic logistic-harvest model without Allee effect with $\lambda = 0.2$, $\epsilon=0.2$ and different values of $x_0$. We use $f(X_j)=X_j[r(1-\frac{X_j}{K})-\lambda]$ and $g(X_j)= \epsilon X_j$ in equation (\ref{EM}). When $x_0 \in (0,K)$ and $x_0>K$, the population approaches to its maximum size. In both deterministic and stochastic models the behaviour of the solution of the models are almost similar. In other words, for any positive initial value $x_0$, $X_t$ goes to $K_1$ as $t\rightarrow\infty$.

The analysis of the stationary density of model (\ref{log-SDE}) which varies under a proportional increase in Gaussian noise $\epsilon$ and $\lambda$ is drawn in Figure \ref{pdf} for $0<\frac{\epsilon}{\sqrt{r-\lambda}} < 2$ and $r=1$. In this case we have noise-induced transition. In Fig. \ref{pdf}a the value of $\lambda$ is 0.75 and  $\epsilon=0.125, 0.25, 0.375, 0.5$. This tells us that proportional increase in linear multiplicative noise can qualitatively change the behavior of the system. When $\epsilon$ becomes smaller and smaller then stationary density $p(y)$ becomes more and more acute, and the maximum point $y_{max}$ of $p(y)$ tends to one. From Fig. \ref{pdf}b we can see that the probabilistic qualitative behaviour of the stationary densities is similar with Fig. \ref{pdf}a with fixed $\epsilon=0.2$ but for different value of harvest rate $\lambda$ ($\lambda=0.36, 0.84, 0.93, 0.96$). Here also for small value of $\epsilon$ and $\lambda$ ( if both go zero), then stationary density $p(y)$ becomes more acute, and the maximum point of $p(y)$ tends to $y_{max} = 1$. If $\frac{\epsilon}{\sqrt{r-\lambda}} \geq \sqrt{2}$, the population dynamic system in (\ref{log-SDE}) has no stationary densities. This shows that the solutions converge to zero (extinction). From this graph we know that the values of the extrema of stationary density depend on the noise intensity $\epsilon$ and harvest rate $\lambda$.
\subsection{ \textbf{Numerical results and biological implications of logistic-harvest model with Allee effect}}
We fixed the value of the parameters $r=1$, $S=1$, $K=3$. Figure \ref{logDet} plots about the phase line diagram of model (\ref{log-h}) and (\ref{Allee-h}), and potential function $V$ of the model (\ref{Allee-h}). Fig. \ref{logDet}a shows us deterministic model (\ref{Allee-h}) has three equilibrium solution at $X_t=0$ ( stable), $X_t=Y_2$ (unstable) and $X_t=Y_3$ ( stable). In Fig. \ref{logDet}b, the area below $Y_m$ is an absorbing zone. For fixed $\lambda=0.2$, Fig. \ref{logDet}c shows model (\ref{log-h}) is always positive while model (\ref{Allee-h}) is negative when the population $X_t < S.$  The phase line diagram of model (\ref{Allee-h}) is plotted in Fig. \ref{logDet}d for $\lambda$ vary (or $0 < \lambda < m_1).$ We observe that the value of $Y_3$ is smaller than the value of $K$, but $Y_2>S$.

Figure \ref{max} plots the harvest yield $\lambda Y_3(\lambda)$  versus eﬀort $\lambda$ and separation $Y_3(\lambda)-Y_2(\lambda)$ versus $\lambda$. From this Figure, we obtain maximum yield when $\lambda = 2.18$, $r=0.1$ and $\beta=100$. For the value of $\lambda$, the separation between the two equilibrium solutions is $Y_3(\lambda) -Y_2(\lambda) = 0.3283$.

The numerical simulation of stochastic logistic-harvest with Allee effect model is given in Figure \ref{simulation}. Fig. \ref{simulation}a shows the solution of (\ref{LGBM2}) with fixed value harvest rate $\lambda=0.2$, $\epsilon=0$ ( no noise) and initial value $x_0$ vary. In Fig. \ref{simulation}b plots the numerical solution of (\ref{LGBM2}) with noise ($\epsilon=0.02$) and $\lambda = 0.2 $ and $x_0$ vary. Clearly seen that the population extinct occurs when $x_0 < S$. While for $x_0 > S$ and $x_0 >K$, the population size approaches to its maximum size $K$.

The graph in Figure \ref{qr} presents the quasi-stationary density of model (\ref{LGBM2}). Here $p(y) \rightarrow 1$ as $L\rightarrow\infty$. In other words the probability of reaching 0 (lower size) before reaching $L$ (maximum size), when considered as a function of initial population size, $p(y)$  \cite{B. Dennis(2002)} has an inﬂection point at deterministic unstable equilibrium $Y_2$. In this Figure inﬂection point is $Y_m$.
\section{Conclusion}\label{Sec6}
We have studied the logistic-harvest model with and without Allee effect driven by multiplicative Gaussian noise.   For the stochastic logistic-harvest model without Allee effect   we obtained exact   solution,  but for stochastic logistic-harvest model with Allee effect  we proved the stability of the solution process. We analyzed the stationary density and the probability of reaching a large population size before reaching a small one, for  the stochastic models (\ref{log-SDE}) and (\ref{Allee-SDE}).

Our numerical experiments demonstrated that the stochastic models in population growth are different from the deterministic models. The main result of our study is that the stochastic model under Gaussian noise perturbation is asymptotically stable. This matches with an important result of the fishery theory.

In the case of the logistic-harvest model without Allee effect, when the harvesting rate $\lambda$ is less than the growth rate $r$, we observe that there exists two equilibrium solutions $X_u=0$ and $X_s$, which are less than the carrying capacity of the population $K$. However, if the harvesting rate $\lambda$ is greater than $r$ (overfishing), there is no fixed point at all and therefore no equilibrium solution.

In the case of the logistic-harvest model with the Allee effect, if  the harvesting rate $\lambda$ is equal to the critical threshold  $m_1=r\left(\frac{(1+\beta)}{4\beta}-1\right)$, we find that there exists only one nonzero equilibrium state. This equilibrium population is less than the carry capacity $K$. However, if the harvest rate  $\lambda$  is less than $m_1$, we have  two positive equilibrium solutions, both the stable and unstable equilibrium solutions are lower than the carrying capacity $K$ and the unstable $Y_2$ is less than the stable equilibrium solution $Y_3$. If the harvesting rate $\lambda$ is greater than the critical threshold $m_1$, there is no fixed point at all and therefore no equilibrium solution.

As far as biological meaning is concerned, we have to catch at a harvest rate $\lambda$ less than the growth rate $r$ in the case of the logistic-harvest model(\ref{log-h}),  and less than the critical point in the case of the logistic-harvest model (\ref{Allee-h}).
\section*{Acknowledgments}
This work was partly supported by the NSFC grants 11801192, 11771449 and 11531006.


\section*{References}
\begin{thebibliography}{00}
\bibitem{A. Friedman(1976)}
 A. Friedman, \emph{Stochastic Differential Equations and  Applications.} Vol. 2, Academic Press San Diego, 1976.
\bibitem {B. Dennis(2002)}
B. Dennis, Allee effects in stochastic populations,\emph{ OIKOS }\, \textbf{96}(389–401), 2002.
\bibitem{B. Dubey}
B. Dubey, P. Chandra, P. Sinha.A model for fishery resource with reserve area. Nonlinear Analysis: Real World Applications, \textbf{4}( 625-637), 2003.
\bibitem{B.Oksendal(2003)}
B.Oksendal, \emph{Stochastic Differential Equations: An Introduction with Applications}. Springer, 2003.
\bibitem{C.Braumann(2019)}
C. Braumann, \emph{Introduction to Stochastic Diffential Equations with Application to Modellinf in Biology and Finance.} Willey, 2019.
\bibitem{Fima C2005introduction}
C. Fima Kelebaner, \emph{Introduction to stochastic calculus with applications}, (Imperial College Press) 2nd Ed, London WC2H 9HE, 2005.
\bibitem{Applebaum Book(2009)}
D. Applebaum, \emph{L\'evy Processes and Stochastic Calculus}, 2nd ed,  Cambride Univ. Press, Cambridge, 2009.
\bibitem{D. Higham (2001)}
D. Higham J., An algorithmic introduction to numerical simulation of stochastic differential equations, SIAM Rev. \textbf{43}(525-C546), 2001.
\bibitem{Daqing Jiang(2005)}
D. Jiang, N. Shi, Y. Zhao.Existance, uniquence and global stability of positive solutions to the food-limited population model with random perturbation.Math.Comp. Model \textbf{42}(651-658), 2005.
\bibitem{D. Heath}
D. Heath E. Platen. A Benchmark Approach to Quantitative Finance. SpringerVerlag, Berlin, 2006.
\bibitem{E. Allen}
E. Allen, \emph{Modeling with It\^o Stochastic Differential Equations}. ISBN:-13 978-1-4020-5952-0(HB), 2007.
\bibitem{F.Braue(2011)}
F. Braue, C. C. Chavez,\emph{ Mathematical models in population boilogy and epidemicology}, 2nd Ed. Springer, 2011.
\bibitem{Allee(2008)}
F. Courchamp, L.Berec, J. Gascoigne,\emph{ Allee Effects in Ecology and Conservation,} Oxford Univ, 2008.
\bibitem{Grimmett}
G. Grimmett R., D. Stirzaker R., \emph{Probability and Random Processes}. 3rd Ed. Oxford: Oxford Univ. Press, 2006.
\bibitem{H. Kinfe(2016)}
H. Kinfe, T. Zebebe, Mathematical modeling of Fish Resources Harvesting with Predator at Maximum Sustainable Yield. \emph{Math. Theory Modeling}.\textbf{6}(2224-5804), 2016.
\bibitem{duan2015introduction}
J. Duan \emph{An Introduction to Stochastic Dynamics}. (Cambridge Univ. Press), 2015.
\bibitem{Drake.Lodge(2006)}
J. Drake M.,D. Lodge M., Allee effects, propagule pressure and the probability of establishment: risk analysis for biological invasions. \emph{Biological Invasions}, \textbf{8 }(365-75), 2006.
\bibitem{D.M.JamesDifferentialDynamicSytem(2007)}
J. Meiss D.,\emph{ Differential Dynamical Systems}, Society for Industrial and App. Math. Philadelphia,PA,USA, 2007.
\bibitem{J.Murray(2002)}
J. Murray D., \emph{Mathematical Biology. An Introduction}, Volume I, Springer Verlag, New York, 2002.
\bibitem{Arnold(2003)}
L. Arnold, \emph{Random Dynamic Systems},Springer,Germany, 2003.
\bibitem{L. Arnold(1972)}
L. Arnold, \emph{Stochastic Differential Equations: Theory and Applications}, Wiley, New York, 1972.
\bibitem{Ricciardi}
L. Ricciardi M., Stochastic Population Theory: Birth and Death Processes. In: Hallam T. G., Levin S. A. (eds) Mathematical Ecology. Biomathematics, vol 17. Springer, Berlin, Heidelberg, 1986.
\bibitem{M. Hasanbulli}
M. Hasanbulli, P. Svitlana Rogovchenko, V.Yuriy Rogovchenko,Dynamics of a Single Species in a Fluctuating Environment under Periodic Yield Harvesting.\emph{ J. Appl. Math.} Article ID 167671, 2013.
\bibitem{panik2014growth}
M. Panik J., \emph{Growth Curve Modeling: (Theory and Application)}. ISBN:978-1-118-76404-6, 2014.
\bibitem{Michael J.Panik(2017)}
M. Panik J., \emph{Stochastic Differential Equations: An Introduction with Applications in Population Dynamics Modeling}. Willey USA, 2017.
\bibitem{M. Kot(2011)}
M. Kot, \emph{Elements of Mathematical Ecology}. Cambridge Univ. Press. New York, NY, 2011.
\bibitem{M. Krstic(2010)}
M. Krstic, M. Jovanovic, On stochastic population model with the Allee effect, \emph{Math. and Computer Modelling} \textbf{52 }(370-379), 2010.
\bibitem{M. Rahmani(2015)}
M. Rahmani H. Doust, The logistic modelling population; having harvesting factor. Yugoslav \emph{J. of Operations Research}, \textbf{25}(107-115)
DOI:10.2298/YJOR130515038R, 2015.
\bibitem{N.F.Britton}
N. Britton F.,\emph{ Essential Mathematical Biology (Springer undergraduate)}.Univ. of Bath, Claverton Down, Bath BA2 7AY, UK. mathematics series, 2003.
\bibitem{N. Saber(2009)}
N. Saber Elaydi, J.Robert Sacker, Population model with Allee Effect: A new model, \emph{J.Bio.Dyam.Vol}.\textbf{00}, No.\textbf{00}, 2009.
\bibitem{P.E.Kloeden}
P. E. Kloeden, E.Platen, \emph{Numerical Solution of Stochastic Differential equations.}Springer Science \& Business Media, 2013.
\bibitem{P. Amarasekare}
P. Amarasekare, Allee Effects in Metapopulation Dynamics.vol. \textbf{152}, no. \textbf{2} \emph{American naturalist}, 1998.
\bibitem{P. Bhattacharya}
P. Bhattacharya, S. Paul, K. S.Choudhury, Maximum Sustainable Yield Policy in PreyPredator System-A Study.2nd International conference on Computing Communication and Sensor Network, 2013.
\bibitem{Qingshan Yang(2011)}
Q. Yang, D. Jiang, A note on asymptotic behaviors of stochastic population model with Allee effect. \emph{App.Math Model}.\textbf{35}(4611-4619), 2011.
\bibitem{S. Lynch}
S. Lynch, \emph{Dynamical Systems with Applicatin using MATLAB}, 2004 .
\bibitem{Simon(2019)}
S. Simo , S. Arno, \emph{Applied Stochastic Differential Equations}.Cambridge Univ. Press, 2019.
\bibitem{S.Pinheniero(2018)}
S. Pinheiro, A Stochastic Logistic Growth Model with Predation: An Overview of the Dynamics and Optimal Harvesting. In: Pinto A., Zilberman D. (eds) Modeling, Dynamics, Optimization and Bioeconomics III. Springer, (313-330), 2016.
\bibitem{S.Pinheiro(2018)}
S. Pinheiro, Optimal harvesting for a logistic growth model with predation and a constant elasticity of variance,\emph{ Ann Oper Res} \textbf{260}(461-480), 2018.
\bibitem{Almaz T.(2020)}
T. Almaz, T. Daniel,K. Anas, J. Brannan, Mean exit time and escape probability for the stochastic logistic growth model with multiplicative $\alpha-$ stable L\'evy noise. https://arxiv.org/abs/2008.00160v1, 2020.
\bibitem{T. Kumar(2010)}
T. Kumar Kar, K. Chakraborty, (2010). Effort Dynamics in a Prey-Predator Model with Harvesting. \emph{International Journal of Information and Systems Sciences,} \textbf{6(3) }(318-332), 2010.
\bibitem{V.M(2011)}
V. Mackeri$\breve{c}$ius, \emph{Introduction to Stochastic Analysis: Integral and Differential Equation.} WILEY, 2011.
\bibitem{X.Mao(2002)}
X. Mao,E. Renshaw, G. Marion, Environmental Brownian noise suppresses explosions in population dynamic, \emph{Stochastic Process.Appl}. \textbf{97}(95-110), 2002.
\end{thebibliography}
\end{document}